\documentclass[12pt,preprint]{aastex}

\usepackage{amssymb}
\usepackage{amsmath}
\usepackage{graphicx}

\newcommand{\mj}   {$J_{550}/I_0$}

\newcommand{\cl}   {$A$}
\newcommand{\clt}  {$B$}
\newcommand{\clre} {$C$}

\newcommand{\tg}   {T_{g}}

\begin{document}

\title{Dust heating by the interstellar radiation field in models of turbulent molecular clouds}

\author{Thomas J. Bethell\altaffilmark{1}, Ellen G. Zweibel\altaffilmark{1,3},
Fabian Heitsch\altaffilmark{2} and J.S. Mathis\altaffilmark{1}}
%\author{\today}
\altaffiltext{1}{Department of Astronomy, University of Wisconsin, Madison}
\altaffiltext{2}{Institute for Astronomy \& Astrophysics, Munich, Germany} 
\altaffiltext{3}{Also Department of Physics, University of Wisconsin, Madison}
%\shorttitle{}
\shortauthors{Bethell et al.}
\begin{abstract}
We have calculated the radiation field, dust grain temperatures, and far
infrared emissivity of  numerical models of turbulent molecular clouds. When 
compared to a uniform cloud of the same mean optical depth, most of the volume inside
the turbulent cloud is brighter, but most of the mass is darker. There is little
mean attenuation from center to edge, and clumping causes the radiation 
field to be somewhat bluer. There is also a large dispersion, typically by a
few orders of magnitude, of all quantities relative to their means.  However, despite
the scatter, the 850$\mu$m emission maps are well correlated with surface
density. 
The fraction of mass as a function of intensity can be reproduced
by a simple hierarchical model of density structure. 
\end{abstract}
%\section{Introduction}
\section{INTRODUCTION}
Observations of molecular clouds indicate that they are inhomogeneous in their      
internal structures. Although                                                      
the mean extinction through such a cloud may be quite high, the extinction along    
selected lines of sight                                                             
can be quite low. The result is that even dense gas in a starless cloud can find itself brightly        
illuminated by the ambient                                                 
Galactic radiation field, with            
consequences for grain                                                              
heating, ionization balance, and photochemistry (e.g. Spitzer 1978, Tielens \& Hollenbach 1985) within the cloud. Although astrophysicists have long been aware that extinction in clouds is highly   
variable - Chandrasekhar \&                                                         
M\"unch (1950) proposed that the statistics of extinction could be used  as a probe 
of interstellar turbulence -                                                        
the relative paucity of detailed cloud models has made it infeasible to explore     
these effects and their observational                                               
consequences. Only in the past few 
years have  dynamical models of                                                     
turbulent molecular gas, based on numerical simulations, become available and some studies of 
extinction and radiative transport in the context of these models have been carried out 
(Padoan \& Nordlund 1999, Juvela \& Padoan 2003).  In this paper we use such models to explore the range of mean intensities, $J_\lambda$, in clumpy clouds exposed to an interstellar radition field (ISRF) appropriate to the solar neighborhood.  Although the principle application of our results is to show the effects of a clumpy gas distribution on grain temperatures and far-infrared emissivity, the general problem of how radiation penetrates a clumpy, dusty medium is important in other situations in astrophysics such as protostellar disks (Wolf, Henning \& Stecklum 1999), the dusty interstellar medium (Witt \& Gordon 2000), the Galactic center (Morris \& Serabyn 1996), and AGN (Krolik 1998).  Because dynamical models are not always available or easy to use, we attempt to identify features of our results which can be reproduced by a simpler model, namely a hierarchical (fractal) density distribution.

In \S 2, we introduce the cloud models, comparing their basic physical attributes, as well as
the representative dust mixture and interstellar radiation field used in the calculations. 
In \S 3, we describe the approach                                                     
used to calculate the penetration of an isotropic, monochromatic ambient radiation  
field into a molecular cloud. We                                                    
decribe tests of the numerical technique and consider resolution effects due to the model clouds. 
Section 4 concerns the results of radiative transfer calculations applied to the model clouds.  
We show that there is little                                                        
diminution of mean radiative intensity from the edge of the cloud to its center,    
and that although there is a                                                        
correlation between local gas density and local radiation field, there is a large 
dispersion about the mean.  In principle our results are easily extended to point sources of radiation within the clouds (Natta et al. 1981).  However, lacking a dynamical model which includes star formation self consistently we have chosen not to do so.  In advance of our grain temperature calculations in \S 5 
we consider the distribution of colors in the intracloud radiation field.

Section 5 deals with the detailed calculation of grain                                                           
temperatures, as well as the resuling far-infrared emission,
 for a spectrum of grain sizes exposed to a representation of the        
interstellar radiation field propagated                                             
through the model clouds.           
Following the results of \S 4, we find only a weak correlation between the infrared 
spectrum and local conditions in                                                    
the cloud, although the detailed consideration of grain temperatures in \S 5 leads us to the conclusion 
that the 850$\mu$m continuum surface brightness is very well  
correlated with column density, considerably more
so than maps made at 100$\mu$m.  Although these results are obtained for the particular grain models of Draine \& Lee (1984, hereafter 'DL84') we believe they are qualitatively robust.
 Section 6 is a summary and discussion.
\section{THE MODEL CLOUDS}
\subsection{The density structure}
The cloud models are based on 3D  simulations of driven MHD-turbulence in a 
cube with periodic boundary conditions, modeling a fraction of the interior of
an isothermal molecular cloud. Table 1 lists the model types
used for this study.
The isothermal equation of state renders the system scale free.
All models start with a cubical structure with a uniform density
and a uniform magnetic field parallel to the $z$ direction.

In all these simulations, the turbulence is driven at the
largest possible scales, i.e. between the wavenumbers $1$ and $2$, at a constant energy input rate.
The driving mechanism is explained by Mac Low (1999).
The code then evolves a self-consistent turbulent cascade, mimicking the response of the ISM to
 turbulent energy input 
at the largest scales.

In the following, we define 
the cloud as the largest possible sphere that fits into the original cubical simulation
domain. This is a somewhat artifical definition, in that it does not respect
the ``natural" structure a cloud might be expected to have; there is no
mean density gradient, and occasionally a dense blob is sheared off at the
edge.  The periodicity of the simulation domain allows for free translation of the
density continuum without the introduction of discontinuities. We have exploited this feature in creating the illustrative brightness maps in \S 5.

The length of the cubical domain is $L=2$. In the subsequent analysis, we will find it
convenient to normalize the models such that the optical depth is fixed. 
We have chosen to scale the density such that the mean center-to-edge 
optical depth (absorption + scattering) $<\tau_{cen}>$ at $\lambda=550$nm 
equals 10, and the radius $R=5$pc, ballpark values for molecular clouds.  
For Cloud $A$ this implies a mean density of $<n_H>=1.3$ $10^3$cm$^{-3}$, corresponding
to a cloud mass $M_{cl}=1.6$ $10^4 M_{\odot}$.  In contrast a uniform cloud 
similarly calibrated requires $<n_H>=1.1$ $10^{3}$cm$^{-3}$ 
($M_{cl}=1.4$ $10^4 M_{\odot}$).

Our measurements begin at system time $t$ = 0.0, when the model has reached an  
equilibrium state between the energy dissipation rate due to shock interaction 
and numerical diffusion, and the driving energy input rate.  The final turbulent 
density continuum has a density range of four orders of magnitude; two orders 
of magnitude above and below the mean. Surface density maps of the MHD simulation used to construct Cloud \cl \space looking along the three coordinate axes are shown in Figure 1.  Also shown are 
histograms of the column densities $N_H$ for the maps normalised arbitrarily to an average column 
of $<N_H>=10^{22}$cm$^{-2}$.  Viewing along the B field (along $z$) reveals no 
strong features that distinguish it from views made perpendicularly to the B 
field (along $x$ and $y$). This is unsurprising, given the relatively weak magnetic field in this
model.

%\input{tab1.tex}

%\begin{figure}
%\label{denmaps}
%\begin{center}
%\plotone{f1.eps}
%\end{center}
%\caption{Surface density maps for model cloud $A$ looking along the three principal axes (\textit{top row}).  Histograms for the column density $N_H=\int n_H ds$ are shown (\textit{bottom row}) for each viewing direction.  The maps each have a mean column density $<N_H>=10^{22}$ cm$^{-2}$.} 
%\end{figure}

\subsection{The ISRF and dust properties}

We bathe the models in the ISRF of Mathis, Mezger and Panagia (1983).
Our sampling of ISRF wavelengths ($\sim 20$ distributed logarithmically in the range $[91$nm$,2\mu$m$]$) ensures that the
attenuated intracloud radiation field $J_{\lambda}$ possesses an
agreeably smooth spectrum when interpolated to unsampled wavelengths.  One needs a more comprehensive sampling of mid-infrared wavelengths to capture the spectral features associated with the polycyclic aromatic hydrocarbons (PAHs), increasing the computational time considerably.  Juvela and Padoan (2003) have addressed this problem by devising a ``library method'',
inferring the overall spectral form from radiative transfer calculations made at a small number of reference wavelengths.  We do not consider PAHs here, since their contribution to the far-IR emission is negligible.

To illustrate the qualitative effects of the intracloud radiation field on dust grains we have constructed a grain ensemble by applying the Mathis, Rumpl \& Nordsieck (1977, hereafter `MRN') grain-size distribution to the graphite and astronomical silicate grains advocated in DL84\footnote{For tabulated optical properties see the extremely helpful website
www.astro.princeton.edu/~draine.}.  This choice of grain ensemble is meant only to illustrate possible effects.  The exact nature of real interstellar dust is unknown despite being the subject of considerable study and speculation (e.g Aannestad 1975; Wright 1987;  Mathis \& Whiffen 1989; Smith, Sellgren \& Brooke 1993). However, the DL84-MRN combination of grain composition and size distribution reproduces most aspects of the mean Galactic extinction curve (Savage \& Mathis 1979).

The scattering phase function was taken to be the Henyey-Greenstein (1941) $\Phi(\omega,g)$ with the albedo, $\omega$, and $g\equiv <\cos\theta>$, the angle of scattering.  Draine (2003) has proposed  another scattering
function, which differs from the HG function by less than 10\% over the
range $0.48\mu$m$<\lambda<0.96\mu$m and is presumably more realistic. We 
stayed with the HG function because it can be manipulated analytically with greater ease.

We calculate temperatures for grains with 
radii distributed logarithmically in the range $0.005\le a \le 0.25 \mu$m.
This grain size range was sampled
densely enough (usually 16 different radii) to ensure that all changes in grain 
temperature across the range were captured.  At the very smallest radii $a < 0.005 \mu$m transient heating may be important (Draine \& Li 2001 and references therein). These very small grains must be treated outside the radiative equilibrium approximation, considering the effects of the individual photon absorption events that momentarily heat the grains to high temperatures.

\section{Dilution of the Ambient Galactic Radiation Field within a Clumpy Sphere}
\subsection{Radiative transfer approach}

We calculate the scattering within the cloud to obtain the mean specific intensity $J_{\lambda}({\bf x})$ at the point ${\bf x}$ by a Monte Carlo approach described in the Appendix, considering only wavelengths for which emission from grains is negligible.  Our technique is a variant of the usual 
Monte Carlo method, which enables us to obtain a relatively high degree of accuracy at a respectable computational cost.  Rather than compute $J_{\lambda}$ throughout the cloud by propagating photons inward from the edge, we chose a sample of interior points at which
we wished to compute the radiation field, selected a sample of incoming ray 
directions, and propagated the rays out \textit{backwards} to the cloud edge (see Lu \& Hsu 2003 and references therein for a discussion of the method in the context of engineering problems).  This 
\textit{reverse} method is a computationally efficient way to calculate the mean intensity 
accurately at a modest but sufficient subsample of points within the model clouds.  
It is particularly effective when applied to self-shielded locations through which relatively few 
photons pass according to most other forward method Monte Carlo schemes.

\subsection{Calculations with a Uniform Cloud}

We establish the accuracy of our radiative transfer code by evaluating the 
mean intensity for the case of a uniformly dense sphere of central optical 
depth $<\tau_{cen}>=10$, and comparing with the 
semi-analytical solutions of Flannery, Roberge \& Rybicki (1980). In much of
what follows, we will use the  
uniform cloud as a standard against which to compare similarly calibrated 
clumpy model clouds.

%\begin{figure}[t]
%\begin{center}
%\plotone{f2.eps}
%\end{center}
%\label{test}
%\caption{\label{fig:test} The relative mean intensity $J/I_0$ computed with the code (\textit{crosses}) at radial positions $0<r<R$ inside a uniform cloud of central optical depth $\tau=20$ and radius $R$.  The solid line is given by the closed-form asymptotic solution for optically thick clouds derived in Flannery, Roberge \& Rybicki (1980).}
%\end{figure}

For each position of interest the radiative transfer code follows rays distributed uniformly in $N$ directions.  For each direction we initiate $M$ rays. 
In its less-than-optimal configuration ($N=16\%$ of the optimum number $N_{opt}\sim 
5\times 10^4$ of rays, $M\sim 10$; see Appendix),
the code achieves a rms error of 1.4\% (see Fig. 2). The small
deviation of the numerical from the analytical solution near the
edge can be traced to the  breakdown of the asymptotic approximation in the analytic solution which
leads to an underestimation of
$J/I_0$. We regard this degree of accuracy as acceptable; it can be improved
by computing more raypaths, but requires more resources. With $N/N_{opt}=
0.16$, $M=10$, 
 $J$ may be found for $4$ $10^4$ points inside a 128$^3$ cell model in about
 one hour on a desktop computer such as a single processor SGI or Sun workstation. All subsequent 
calculations use the code 
in this configuration unless noted otherwise.

\subsection{Effects of model resolution on radiative transfer}

With clumped models the geometric center of the cloud has only one unique property:  at a reference wavelength ($\lambda=550$nm) we specify the average central optical depth within the cloud by computing the column density along radial paths from the center to the edge of the cloud.  We force their average optical depth to be the specified value (usually 10).  Every other wavelength has a mean central optical depth in proportion to the dust opacity relative to $\lambda=550$nm.

In this section we discuss only the results for $\lambda=550$nm, in which the mean central radial optical depth is 10.

A major goal is to determine the effects of clumping on the distribution of mass with $J_{550}$.  For convenience in plotting, we define the mass distribution function $df_m/d\log_{10}(J/I_0)$ to be the fraction of the cloud mass per unit $\log_{10}(J/I_0)$.  We also consider a similar distribution function $f_V$ for the volume fraction.

We first consider the effects of spatial resolution on the distribution functions.  Will small, dense, dark clumps that are unresolved in coarse-grained simulations contain appreciable amounts of mass at small values of $J$?   
The simulation cubes are available (see Table 1) at resolutions of
$512^3$, $256^3$ and $128^3$ cells, the 
highest resolution simulations possessing small scale structures that 
cannot be resolved by runs at the lower resolutions. These extra structures 
potentially provide windows through which radiation can stream, with less 
attenuation, increasing the mean intensity within the cloud. On the other
hand, at increased resolution it is possible to form tiny, dense cores
which potentially are extremely dark. It is of interest to see how the
resolution of the models affects our computations of $J_{\lambda}({\bf x})$ and the distribution functions.

Models with the same input parameters but run at different
resolution are actually dynamically different from each other, because the 
random driving pattern used to drive the turbulence will take on different 
realizations with varying resolution, as will the numerical diffusivities. Therefore,
we compare a sequence of models derived from the 512$^3$ model by
degrading the resolution.
A $512^3$ cell cloud may be smoothed into a $256^3$ cell cloud by taking a cube of eight cells
and replacing them by one supercell having their average density, thereby conserving the mass.
Repeated smoothings must eventually render the cloud uniform, the result of 
which is typically a darker cloud throughout the volume.  

The effects of smoothing a cloud
from 512$^3$ to lower resolutions (256$^3$, 128$^3$ and 64$^3$) are shown in Figure 3. 
The left panels in Figure 3 shows $J(512^3)/J$ for various other resolutions, plotted against the densities, $n_H$, in the cells.  The 512$^3$ model is model C in Table 1.  The coarsest (64$^3$) resolution is in the lowest panel in the figure.  As the resolution coarsens we see that the range of densities narrows, the spread in $J$ at a particular $n_H$ increases, and the mean $J$ decreases for all densities except those at the high end.  The right panels show the changes in $J$ within individual cells brought on by coarsening the resolution.  At the coarsest grid we considered (64$^3$), the mean $J$ is not only broadened but also significantly \textit{decreased} on the average because the radiation cannot easily penetrate through low-density cells since these have been eliminated by the smoothing process.  Conversely, at the highest densities the smoothing process tends to \textit{increase} the energy 
that penetrates into these regions; the dense, dark cores resolved at a resolution of 512$^3$ 
are smoothed over into less dense and more penetrable regions.  Overall, smoothing the model from a resolution of 512$^3$ to 256$^3$,128$^3$ and 64$^3$ cells \textit{decreases} the total energy (according to a volume average) that penetrates into the cloud by approximately 1\%,6\% and 60\% respectively.    

%\begin{figure}
%\label{fig:res_test}
%\begin{center}
%\plotone{f3.eps}
%\end{center}
%\label{res_test}
%\caption{Comparison of mean intensity at equivalent points (i.e cell coordinates $(x,y,z)$ 
%in the 512$^3$ cell model become $(x/2,y/2,z/2)$ after smoothing to 256$^3$ cells) in model 
%\clre \space smoothed from 512$^3$ to 256$^3$,128$^3$ and 64$^3$ cells.}
%\end{figure}

Since the results at 512$^3$, 256$^3$, and 128$^3$ are quite similar to
each other, while the results at 64$^3$ are noticeably different, it is
tempting to say that the models are converging and that a resolution of
128$^3$ is adequate for determining the distribution functions and other statistical properties of the radiation field. Quantitatively, however, the criterion for
convergence is not entirely clear.
Applying the fiducial optical depth of $\tau \sim20$ across a model of 
resolution $\mathcal{R}$ yields a mean-free path of $\mathcal{R}/20$ cells, 
a size-scale resolved adequately by even the 64$^3$ smoothed model. 
However, there is extreme clumping in the model (about 4 orders of magnitude) on size scales comparable to the radius of the cloud.  The porosity allows radiation to penetrate deeply if the passages are spatially resolved.  The 64$^3$ resolution evidently 
fails to provide sufficient porosity.
\section{CALCULATIONS WITH THE MODELS}

\subsection{Mean Intensity}

In what follows we ran the Monte Carlo radiative transfer code with the model clouds, calculating the mean intensity $J_\lambda$ for a density weighted random sample of $8 \times 10^4$ points.  The density weighted sampling of points is akin to picking hydrogen atoms at random and asking what the ambient radiation field is like in the vicinity of that atom.  Since the density `continuum' is defined on a grid, we do not attempt to define $J_\lambda$ on smaller scales; instead we assume it is uniform within each cell.  However, the radiative transfer method can in principle calculate $J_\lambda$ on arbitrarily small scales. 

Figure 4 shows various distributions that illustrate the basic statistical properties of the intracloud radiation field.  The top panel (Fig. 4a) shows the scatter of $J_{550}/I_0$, the relative mean intensity at $\lambda=550$nm, plotted against density $n_H$.
On average $J_{550}$ drops as $n_H$ increases 
(Fig. 4a) but the scatter about this trend is extremely large.  
The sharp upper limit to \mj\space is readily identified with 
points lying near the cloud's surface ($0.98<\frac{r}{R}<1.0$), illuminated 
by $\sim 2\pi$ steradians of almost unattenuated ISRF in addition to a small contribution 
that passes through the cloud. In 
Figures 4b and 4c we show the 
mass and volume distribution functions respectively.  For comparison the result for 
the uniform cloud is also shown.  The distribution of the model cloud's mass favors lower values of \mj \space 
(mass distribution peaks at \mj$ \sim 0.1$) than in the uniform cloud, 
with about 16\% of the cloud's mass associated with a low intensity tail 
(\mj$<0.01$).  By volume, the model cloud is brighter than the uniform 
cloud.  These two effects result from an overall anticorrelation between $J_{550}$ and $n_H$, suggested in Figure 4a and 
shown explicitly in Figure 4d.  In Figure 5 we show a plot similar to Figure 4a but evaluated at $\lambda=333$nm.  It shows that at shorter wavelengths, at which the cloud is optically thicker, the scatter in $J_\lambda$ is generally larger.  

%\begin{figure}
%\begin{center}
%\plotone{f4.eps}}
%\end{center}
%\caption{\label{fig:bethell_plot}{\bf a} The relative mean intensity \mj \space at $\lambda=550$nm for a random sample of points within cloud $A$, where $I_0$ is the intensity of unattenuated interstellar radiation, assumed to be isotropic. The average \mj \space for a given density $n_H$ is shown by the bold line.  {\bf b} The fraction of the cloud mass per unit $log_{10}($\mj$)$, $df_m/dlog_{10}($\mj$)$. {\bf c} The equivalent volumetric distribution $df_v/dlog_{10}($\mj$)$.  {\bf d} The average overdensity associated with a value of \mj \space (i.e. $df_m/df_v$).}
%\end{figure}

%\begin{figure}
%\begin{center}
%\plotone{f5.eps}
%\end{center}
%\caption{Like Figure 4a but calculated at $\lambda=333$nm.  Note the scatter and compare with Figure 4a.}
%\end{figure}

The intracloud radiation fields at $330$nm (\textit{dashed lines}) and $550$nm (\textit{solid lines}) averaged in thin spherical shells of radius $r$ are shown in Figure \ref{fig:UV_onion}.  The average central optical depths are approximately 16 at 330 nm and 10 at 550nm.
 Compared to the uniform cloud, the mean intensity in the model cloud is 
enhanced dramatically, and is insensitive to radial position in the 
inner 50\% of the cloud's volume ($r<0.8R$), increasing rather sharply near the cloud's 
outer surface ($r>0.8R$). There is also less reddening from
edge to center (we discuss color in \S \ref{color}).  Despite the average 
uniformity of \mj\space in the cloud's interior, it should be recalled that
 the scatter from point to point is large throughout (see Fig. 4a).

%\begin{figure}

%\begin{center}
%\plotone{f6.eps}}
%\end{center}
%\caption{\label{fig:UV_onion} For $\lambda=333$nm (\textit{dashed} line) and 
%$\lambda=550$nm (\textit{solid} line) the mean specific intensity is averaged o%ver the volume of thin spherical shells to form the shell average $<J_{\lambda}>$.  
%The shells are centered on the cloud center, of radii $0<r<R$ and thickness $R/50$, where $R$ is the cloud radius.  
%Results for the uniform cloud with the same central optical depth ($<\tau_{cen}>=10$) are shown for comparison.}
%\end{figure}

\subsection{\label{color} Intracloud Colors}

The color variation
 inside the cloud is the result of the differential spectral extinction by the dust.  The mean intensity $J_{\lambda}$ at each  point inside the cloud is formed from an average over the rays propagating from the point to the surface of the cloud, each ray contributing differently to $J_{\lambda}$ by virtue of their different path histories.  At some other wavelength the optical properties of the scattering dust will be different, and so too the transfer of radiation which must be recalculated for each new wavelength.  As a result one might expect a scatter in colors amongst points with similar $J$ at some fixed wavelength.  Possible 
degeneracies arise: locations of similar $J_\lambda$ at some wavelength may be bathed in
 $4\pi$ steradians of highly reddened, diffuse glow; a small
 but unattenuated shaft of pristine ISRF; or, most plausibly, some 
intermediate case.

%\begin{figure}[t]
%\label{fig:color_scatter}
%\begin{center}
%\plotone{f7.eps}
%\end{center}
%\caption{The color $J_{333}/J_{550}$ for points inside model \cl.  For comparison the colors for two uniform clouds are shown: Uniform 1 is calibrated to $\tau_{cen}=10$ whereas Uniform 2 has a total mass equal to that of model \cl.}
%\end{figure}

Figure 7 illustrates the effect of clumpiness on
intracloud colors; the dark locations typically see the most reddened
radiation fields, but there is a spread of colors for a given $J_{550}$.
In light of the anti-correlation between mean intensity $J$ and density $n_H$ one recovers the expected result that dense places are typically reddened. Nevertheless, over the innermost $50\%$ of the cloud's volume the average color seen by dust grains is considerably bluer (up to a factor of $100$ for $J_{333}/J_{550}$) and less dependent on radial location than the color in uniform clouds with either the same optical depth or same mass (Fig. 8).

%\begin{figure}
%\begin{center}
%\plotone{f8.eps}
%\end{center}
%\caption{\label{fig:color_onion}The color $J_{333}/J_{550}$ averaged in spherical shells 
%centered in Cloud \cl.  The averaging is density rather than volume weighted, reflecting 
%the average color seen by an H atom or dust grain in these shells.  The error bars represent 
%the \textit{rms} scatter about the mean.  For comparison two uniform cloud results are shown; 
%Uniform 1 with $<\tau_{cen}=10>$ and Uniform 2 with the same mass as model \cl.}
%\end{figure}

The color and more generally the spectral shape of the radiation field is an important factor 
in grain heating (see \S \ref{graint}), and must also play
a role in cloud chemistry. 

\subsection{Comparison between Fractal and Turbulent MHD Clouds}

The turbulent cloud models used in this study were originally generated to
study cloud dynamics, and were developed at considerable
computational cost. As we discussed in \S\S 2 and 3.3, they suffer from
their own idealizations: finite numerical resolution and a probably
unrealistic form of dynamical driving are two of them. In order to probe 
which features of
the model are robust, and investigate whether a simpler prescription for generating 
density structure might give similar results, we investigated the
radiation field in a hierarchical model of the gas density which is
intended to replicate a fractal density distribution (Elmegreen 1997).
 
The fractal clouds are grown from seeds, uniquely determining the precise 
density structure for a given fractal dimension $D$.  
The clouds are grown from an initial casting of 32 points according to the 
seed and fractal dimension $D$.  
In each of the subsequent castings a further 32 points are cast about each 
extant point according to the fractal dimension.  
In a total of four castings a total of $32^4$ points are therefore cast, and if 
rendered on a $64^3$ cell grid the average density is $4$ points per 
cell.  In the instances where a uniform background is desired a further 
$2$ points per cell are added.  Finally the fractal clouds are calibrated to $<\tau_{cen}>=10$.    

%\begin{figure}
%\label{fig:every_bethell_plot}
%\begin{center}
%\plotone{f9.eps}
%\end{center}
%\caption{Comparison between, uniform, fractal (dimension D, with and without backgrounds 'b.g') 
%and model clouds \cl \space and \clt .  In the case of the fractal clouds average results have 
%been shown with standard deviations (large error bars), generated by averaging f$_m$ for a number 
%of clouds with the same physical parameters but grown from different seeds.  The model clouds' 
%results are time averages during their steady-state phases (during which there are three time dumps); the deviations (small error bars) 
%are then to be interpreted as intrinsic time variations rather than the differences between clouds sharing the same physical parameters but realised separately.}
%\end{figure}

Figure 8 shows $df_m/d\log_{10}($\mj$)$ for both fractal and MHD models.  The heavy solid line is a uniform density model.  There are four types of fractals: fractal dimension 2.3 and 2.6, each containing either no uniform density background or 1/3 of the mass in such a background.  For each type 
of fractal distribution six models were calculated, each with a different initial seed that determines the locations of the 32 points in the first casting of points.  The final model contains hierarchically clumped clouds around each of the initially placed points. 

The ``error bars'' in Figure 9 are \textit{not} errors, but are the upper and lower envelope of the six individual mass distributions for the fractal dimension $D=2.6$ and no background density.  The deviations of the other fractal distributions show similar deviations.  The smaller ''error bars'' around the light solid line shows the time variability in the MHD models.

The large scatter in the fractal results indicates that the cloud structure 
and radiation transport properties depend strongly upon the seed from 
which the cloud is grown.  This scatter means that it is difficult to differentiate 
between fractal dimensions (either with or without uniform backgrounds) based upon 
their $f_m$ distributions.  The fractal models with a background density 
contain a greater fraction of mass at intensities brighter than $\log_{10}($\mj$)=-0.5 $
than do similar fractal models without a background. 
 The addition of a background to the fractal clouds generally produces average $f_m$ 
distributions similar to those for the model clouds. Therefore, these
fractal models appear to be a promising alternative testbed for studying
the properties of radiation in clumpy clouds. At the same time, these
results imbue the dynamical models with a pleasing degree of generality.  
  
The relatively small characteristic scatter amongst the MHD distribution functions corresponds to 
the use of the different time dumps in lieu of multiple MHD clouds rendered separately but sharing 
the same physical parameters.  This time difference is so small (about 0.2 dynamical crossing
times) that the MHD models are physically correlated on large scales. Note that even with this
relatively small time variation, the curves for models A and B - the former gas dominant, the latter
magnetically dominant  - are nearly indistinguishable. 

%\section{Grain Temperatures and Emissivities in Clouds heated by the ISRF}
\section{GRAIN TEMPERATURES AND EMISSIVITIES IN THE MODEL CLOUDS}
\label{graint}

In Section \S 4 we investigated the extent to which monochromatic radiation
at selected wavelengths 
penetrates our model clouds.  The calculation of dust grain temperature is a matter of determining the spectral energy density throughout the cloud and 
the subsequent selective absorption and re-emission by the grains themselves.  It should be 
emphasised that the choice of grain model has been made merely to illustrate possible grain heating 
effects, and that the resultant emissivities are not intended to be firm predictions.  
    
\subsection{Basic Equations.  Calculating $\tg$ }

The steady state equilibrium temperature, $\tg$ of grains bathed by a mean 
intensity 
$J_\lambda$ is determined by balancing the radiative energy absorbed
with that emitted thermally (see DL84),
\begin{equation}
\label{eqn:energybalance}
\int_{0}^{\infty} Q_{abs}(a,\lambda) J_\lambda d\lambda = \int_{0}^{\infty} Q_{abs}(a,\lambda) B_{\lambda}(T_g) d\lambda,
\end{equation}
where $Q_{abs}(a,\lambda)$ is the absorption efficiency for a grain of radius $a$ at wavelength $\lambda$ and $B_{\lambda}(T_g)$ is the blackbody function evaluated at the dust grain temperature $T_g$.

For very small grains ($a<0.01 \mu$m) the steady state
approximation begins to break down;
individual UV photons are sufficiently energetic to heat small grains to
relatively high temperatures which subsequently cool between absorption events. 
These transiently heated grains primarily re-radiate their energy shortward of 
$\lambda=100\mu$m, and may be safely excluded in this FIR analysis; for a treatment of sporadic heating of small grains in clumpy clouds see Juvela \& Padoan (2003).

\subsection{Results}

%\begin{figure}[t]
%\label{fig:ISRFtemps}
%\begin{center}
%\plotone{f10.eps}
%\end{center}
%\caption{Temperatures for graphite and silicate grains of radii $a$ exposed to the unattenuated ISRF.  The equivalent DL84 graphite (\textit{diamonds}) and silicate (\textit{triangles}) temperatures are shown for comparison.} 
%\end{figure}

%The temperature of grains bathed by the unattenuated ISRF are shown in 
%Figure \ref{fig:ISRFtemps} and are in satisfactory agreement with those 
%presented in DL84.

Qualitatively, the temperature of a grain is determined by the contrast between the grain's absorption efficiencies in the visible and FIR spectral regimes.  It is often stated that small grains are hotter than 
large grains because they are to radiate away their energy at wavelengths much larger than their radii.  In order to reach a radiative equilibrium they must attain high internal temperatures, and 
corresponding black-body radiation densities peaked at shorter wavelengths.  This reasoning assumes that all grains absorb a similar 
amount of radiation per unit grain area.  Since the absorption 
efficiency is of order unity when $a>\lambda$, this assumption is a fair approximation 
for large grains absorbing the ISRF.  However, when the radiation field is somewhat reddened, and relatively more
photons satisfy $a<\lambda$, an opposing effect
comes into play.  Because the grain absorption cross section reaches a maximum when $a\sim \lambda$, the reddened radiation field preferentially heats the large grains. This
tends to reduce the temperature discrepancy between grains of different 
sizes.

 The mean temperatures attained by graphite and silicate grains of various radii $a$ are shown in 
Table 2 for model \cl.  As mentioned above,
transient heating is important for the smallest grains; we include them
here merely to make the point that there is now little variation in temperature with grain size.
Table 2 also gives the \textit{rms} variation in temperatures amongst 
similar grains and shows it to be a decreasing function of grain size.  This occurs because of the declining importance of the highly variable blue part of the intracloud radiation field.

%\input{tab2.tex}

%\begin{figure}
%\label{temp_bethell}
%\begin{center}
%\plotone{f11.eps}
%\end{center}
%\caption{The mass distribution function for Cloud $A$ (\textit{solid} line), such that $\int (df_m/dT) dT = 1$.  The mass includes graphite and silicate grains that fall within the MRN size range $[0.005\mu$m$,0.25\mu$m$]$.  The separate contributions of the graphite and silicate grains are shown by the \textit{dashed} and \textit{dot-dashed} lines respectively.}
%\end{figure}

The fraction of the cloud's dust mass, irrespective of grain type or size, 
found at temperature $T$ is shown in Figure 10.  
The distribution of the mass is weakly peaked at low temperatures, $T\sim 10$K with virtually all the mass contained in the MRN distribution being in the temperature range $6-17$K.

\subsection{Dust Emissivity}

Given a grain temperature $\tg$ (which generally varies from point to point and amongst grain types and sizes) the emissivity contribution $dj_\lambda$ from grains with radii
[$a,a+da$] is given by

\begin{equation}
dj_\lambda=C_{g,s} n_H a^{-3.5} \times 4\pi a^2 \times \pi B_\lambda(\tg) Q_{abs}(a,\lambda) da,
\end{equation}
where $B_\lambda(\tg)$ is the black-body function and $Q_{abs}(a,\lambda)$ the
emission efficiency.

%\begin{figure}
%\begin{center}
%\plotone{f12.eps}
%\end{center}
%\caption{\label{fig:emis_by_lambda_by_a}The differential contribution $dj_\lambda /da$ of grains of
%  different radii to the spectral
%  emissivity $j_\lambda$ at $\lambda=100,350$ and $850$ $\mu$m (\textit{top} to \textit{bottom}).  
%  Each line represents the emissivity of dust associated
%  with a randomly chosen H atom except the \textit{dashed} line which
%  is the emissivity associated with an H atom exposed to the
%  \textit{unattenuated} ISRF (i.e. an upper limit).} 
%\end{figure}

The spectral emissivity of dust grains $dj_\lambda/da$ associated with
a sample of randomly chosen mass is shown in Figure 11. 
Grains of large radii are readily identified as the principal source of the $850\mu$m 
emission.  The smaller grains, unable to radiate away
their energy at long wavelengths,  dominate the emissivity at shorter wavelengths $\lambda\sim100 \mu$m. 

%\begin{figure}
%\begin{center}
%\plotone{f13.eps}
%\end{center}
%\caption{\label{fig:emis_scat} Specific emissivity per H atom $j_{\lambda,H}$ 
%at $\lambda=100,850\mu$m for a random sample of atoms situated in locations of 
%density $n_H$ in model \cl.  The solid lines are the mean values of $j_{\lambda,H}$ 
%associated with density $n_H$. }
%\end{figure}

Since $850\mu$m is well into the Rayleigh-Jeans region ($hc/\lambda kT <<1$) of the spectra, the small 
temperature variations of the large grains responsible for this emission correspond 
to small variations in the $850\mu$m emissivity.  On the other hand, $100\mu$m lies 
in the Wien spectral regime where even small temperature variations can cause 
large emissivity changes (Fig. 12).  The small grains, sensitive to the highly variable blue 
part of the intracloud radiation field, find themselves with a relatively large 
range in temperatures. The $100\mu$m emission is therefore intrinsically more 
variable than the $850\mu$m emission.  The $850\mu$m emissivity \textit{per H atom} 
$j_{\lambda,H}$ varies far more slowly throughout the cloud, the volumetric 
emissivity $j_{\lambda}$ is then roughly proportional to density $n_H$.

% and brightness maps should correlate well with column density $N_H=\int n_H ds$ ($\mathcal{C}\sim 0.996$) (Figures \ref{fig:emis_scatter} and \ref{fig:brightness_maps}).   $j_{\lambda,H}$; the $100\mu$m brightness map becoming somewhat obfuscated by this intrinsic variability ($\mathcal{C}\sim0.85$).

Brightness maps (scaled so that their maxima equal 1) made by integrating the spectral emissivity $j_\lambda$ ($\lambda=100$ and $850\mu$m) and bolometric emissivity $j_{bol}$ through the cloud, $I_\lambda \propto \int j_{\lambda,H}n_H dx$, are shown alongside the surface density in Figure 13.  The maps were made for a central cubical region ($32^3$ cells), looking along the $x$ axis.  Since the central parts of the cloud possess a radiation field insensitive to radial location (see Fig. 7), the maps show few of the 'edge effects' that result from a close proximity to the imposed spherical cloud surface.  Instead, we primarily observe the effects of clumpiness.  One can see that the brightness map at $850\mu$m corresponds best with the surface density map $N_H$.

%\begin{figure}
%\begin{center}
%\plotone{f14.eps}
%\end{center}
%\caption{\label{fig:brightness_maps} Surface density $N_H$, 100$\mu$m, 850 $\mu$m and bolometric brightness maps (\textit{left} to \textit{right}), scaled such that their maxima equal one.} 
%\end{figure}

Our results have bearing on the problem of directly determining the dust mass corresponding to a FIR/sub mm flux measurement. This problem has received some study, inspiring a number of different approaches
(Hildebrand 1983, Xie et al. 1993; Hobson \& Padman 1994; Li, Goldsmith \& Xie 1999; Xie, Goldsmith \& Zhou 1991). The large scatter in emissivity per H atom at 100 $\mu$m, and
the relatively small scatter at 850 $\mu$m, further suggests that inferring column densities from FIR emission is best done using observations at longer wavelengths.

\section{SUMMARY AND DISCUSSION}

We have studied the attenuation of the interstellar radiation field by dust in models of clumpy
clouds and calculated the dust temperatures and far-IR emissivities.
The main conclusions are as follows: 

1. Inhomogeneity of the density continuum dramatically increases the radiant energy that penetrates 
into the cloud,
 making the \textit{volume} markedly brighter when compared to a uniform cloud of comparable optical depth (Fig. 4c).  The mass contained in clumpy structures provides enough local attenuation that most of the \textit{mass} is in fact associated with lower intensities (Fig. 4b) when compared to the uniform cloud.  

2.  The above two effects imply an anti-correlation of average $J$ with $n_H$, but from point to 
point there exists an extremely large scatter about this trend, more so at wavelengths at which the 
cloud is optically thick (Figs. 4a and 5).  
Despite this large scatter, the mean intensity \textit{averaged} by spherical shells is largely independent of radial position within the inner
50\% of the cloud volume (Fig. 6).

3.  The cloud's inhomogeneity allows small amounts of weakly unattenuated blue ISRF to penetrate the cloud which, when combined with the predominant reddened field, \textit{makes the intracloud field somewhat bluer than in the uniform cloud}.  The variability of this effect introduces a scatter in the color (Fig. 7) which in accordance with $J$ is also on average quite uniform throughout the model (Fig. 8).

4. The differential mass per intensity distribution of the MHD models can be qualitatively
reproduced by fractal clouds generated according to a simple prescription and augmented by a
uniform density background. In both types of inhomogeneous model, the peak of the mass distribution
is shifted occurs at lower intensity than in a uniform cloud, suggesting that this is a robust
feature of clumpiness (Figure 9). These results are relatively insensitive to fractal dimension or
magnetic fieldstrength over the range explored ($D$ = 2.3 - 2.6; $\beta$ = 0.05 - 4.04).

%4.  Attempting to reproduce the effects of inhomogeneity with more tractable cloud models intracloud radiation fields were calculated for fractal clouds 
%generated using a simple prescription.  Although the initial condition (the 
%seed) for growing each of the fractal clouds introduced a considerable 
%scatter, the average results suggested that the fractal dimension $D$ is  
%less important than the presence of a background density in producing 
%results similar to those for the turbulent MHD model clouds (Fig. 
%\ref{fig:every_bethell_plot}).  For some purposes the use of fractal clouds 
%grown using simple rules could act as convenient substitutes for MHD
%models of clouds produced at greater computational expense. This exercise
%also demonstrates that at least some of the trends observed in the MHD
%models are, in fact, robust.

5. For one grain model (DL84) bathed in the attenuated ISRF, the sensitivity of dust grains of 
different radii to the overall spectral form (i.e color) 
of the intracloud radiation field $J_{\lambda}$ produces average grain temperatures 
which do not depend strongly on the grain size
(Table 2).  \textit{Importantly, the smaller grains, absorbing the most variable part of the intracloud radiation field, exhibit a greater scatter in their temperatures}.  When considering how the dust mass is 
distributed with temperature there is a slight preponderance for low temperature ($T\sim 10$K) 
material (Fig. 10), primarily due to the abundance of silicates in the DL84-MRN grain ensemble.

6. Small, relatively hot grains emit the majority of the $100\mu$m emission, larger
 grains emitting longward of this (Fig. 11).  The 
temperature variations in these two grain populations, considered in the 
context of the Rayleigh-Jeans and Wien spectral regimes, yield emissivities 
$j_{\lambda,H}$ exhibiting very different point to point variations, shown in 
Figure 12.  The $850 \mu$m emissivity per H atom varies 
by less than a factor of 5 whereas the $100 \mu$m emissivity per H atom 
varies by several orders of magnitude.
  
7.  In constructing brightness maps (Fig. 13) the intrinsic scatter in $j_{\lambda,H}$, the emissivity per H atom, 
that ultimately arises from the cloud's inhomogeneity increasingly decorrelates maps made 
at $\lambda<100\mu$m when compared with the surface density.  Conversely, maps made at 
longer wavelengths show increasingly higher correlations between brightness and surface 
density maps, and should be preferred if one wishes to infer surface densities 
and cloud masses from brightness maps.

\acknowledgements We are happy to acknowledge support from NSF Grants
AST-0328821 to the University of Wisconsin and AST-9800616 to the University
of Colorado, and the Graduate School of the University of Wisconsin, 
Madison. TJB is grateful for the hospitality of JILA and the Laboratory for
Computational Dynamics at the University of Colorado. FH was supported in
part by a Feodor-Lynen Fellowship from the Alexander von Humboldt Foundation.  The 512$^3$ MHD 
model was made available by Pakshing Li in advance of publication.  We would also like to thank C. 
McKee for his insightful suggestions.  This work was partially supported by National Computational 
Science Alliance (NCSA).
% under {\bf XXXX grant number XXXX and utilized  XXXX alliance system name XXXX.}

\newpage

%%%%%%%%%%%%%%%%%%%%%%%%%
% Here's another way to do the references. I'm not sure whether it's actually better.
% Advantage is that you can get the bibitem-format from ADS directly (so you don't have to type all the authors etc),
% and you can do quotes by \cite{<label>} resulting in (Author Year) or \citet{<label>} resulting in Author (Year).

%\begin{thebibliography}{}
%\bibitem[Brandenburg \& Zweibel(1994)]
%        {BRZ1994}
%        Brandenburg, A. \& Zweibel, E.~G.\ 1994, \apj, 427, L91
%\bibitem[Brandenburg \& Zweibel(1995)]
%        {BRZ1995}
%        Brandenburg, A. \& Zweibel, E.~G.\ 1995, \apj, 448, 734
% .....
%\end{thebibliography}

\newpage
\appendix
\subsection{MONTE CARLO METHOD}

We consider the trajectories of individual photons, each carrying a weight $W$.  The weight reflects the probability of the photon surviving the course of its trajectory without being absorbed (e.g Witt 1977).  The Monte Carlo aspect of the code deals exclusively with the scattering processes, that is, in selecting a probabilistically weighted sample of possible trajectories that connect a point within the cloud to points on the cloud's surface.  It is along such a trajectory that the weight is calculated,
\begin{equation}
W=exp(-\tau_{a}^{tot})
\end{equation}
where $\tau_{a}^{tot}$ is the total absorption optical depth along the trajectory.  The generation of trajectories and their respective weights is sufficient to evaluate the mean intensity. To explore the possible trajectories which connect a particular location inside the cloud to the external radiation field we have implemented the following \textit{reverse} approach for generating photon trajectories;

%\begin{figure}
%\begin{center}
%\input{f14.eps}
%\end{center}
%\caption{\label{diagram} The spherical cloud sits snugly inside the cubic domain of $R^3$ cells.  An observer {\bf A} sends out $M$ rays (one of which is shown) for each of the $N$ directions (${\bf \hat k_{obs}}$) sampling the observe's entire sky. Each ray is evolved according to the Monte Carlo sampling of free paths between scatterings and scattering angles until it reaches the cloud's edge ({\bf B}).  The \textit{dashed} lines are other trajectories joining the observer {\bf A} of photons ($h\nu$) to the ISRF.}
%\end{figure}

An observer is placed at point {\bf A} inside the cloud (Fig. 14) insisting upon knowing the intensity in $N$ directions ${\bf \hat k_{obs}}$ uniformly distributed over the $4\pi$ steradians of sky.  For each of these $N$ directions ${\bf \hat k_{obs}}$ we initiate $M$ \textit{reverse} trajectories.

Proceeding one trajectory at a time the free path between scattering centres, $\bf {x_i}$, is found from probabilistically sampling the possible optical depths  $\tau_{s}$ to the next scattering event.  The cumulative probability $p$ of a photon advancing $\tau_{s}$ before scattering is

\begin{equation}
\label{probtau}
p=exp[-\tau_{s}].
\end{equation}

Rearranging yields

\begin{equation}
\label{mctau}
\tau_{s}=-ln(p),
\end{equation}
where random numbers $p$ sampled uniformly in the interval [0,1] will correctly reproduce the distribution function (Eq. \ref{probtau}).  The random number generator RAN2 (Press et al. 1992) is used to provide a value $p$ which then gives $\tau_{s}$ using equation (\ref{mctau}).  The free path in real space is found by evaluating $\tau^{'}_{s}=\sum \sigma_{s}(\lambda) n_H \Delta l$ in a stepwise fashion through the cloud until $\tau^{'}_{s}$ approaches $\tau_{s}$ to within an acceptable error margin.  

The trajectory is updated.  The free path vector to the next scattering centre is ${\bf l}= \sum \Delta {\bf l_i}$ and the optical depth to pure absorption $\tau_{a}=(\omega^{-1}-1)\tau_{s}$ where $\omega$ is the grain albedo.  The \textit{total} optical depth to pure absorption calculated along the trajectory thus far, $\tau_{a}^{tot}$, is incremented by the amount $\tau_{a}$.

The photon has now reached the location of a new scattering event.  To find the direction ${\bf \hat k_{i}}$ in which the trajectory proceeds, one randomly samples the scattering phase function $\Phi(\theta)$.  The scattering process is direction independent - the reverse scattering process is exactly the same as the forwards process.  For $\Phi(\theta)$ we chose to use the Henyey-Greenstein (HG) phase function (Henyey \& Greenstein 1941);
\begin{equation}
\label{HG}
\Phi(\theta)=\frac{1}{4\pi}(1-g^2)/[(1+g^2-2g\cos \theta)]^\frac{3}{2}.
\end{equation}
where $\theta$ is the trajectory's deflection (polar) angle and the asymmetry parameter $g=<\cos \theta>$ which takes a value between -1 (backwards scattering) and +1 (forwards scattering).  The HG phase function is acceptably accurate ($<10\%$ error) in the range $0.48\mu$m$<\lambda<0.96\mu$m (for the DL84 grains) and can be manipulated analytically with relative ease.  Typical values of g for the interstellar dust from DL84 fall in the range g$\sim0.15$ in the NIR and g$\sim0.65$ in the NUV. 

Integrating to obtain the cumulative probability function for equation (\ref{HG}) and inverting the result yields $\theta (p)$,

\begin{equation}
\label{HGinverted}
\theta (p)=\frac{(1+g^2)-[(1-g^2)/(1-g+2gp)]^2}{2g}.
\end{equation}

As before, $p$ is a random number uniformly distributed in the range $[0,1]$.  Finally, the azimuthal deflection angle is given by $\phi = 2\pi p$ where $p$ is another random number.  Application of the Euler transformations on the pre-scatter direction ${\bf k_{i}}$ and deflection angles $(\theta,\phi)$ yields the new direction ${\bf k_{i+1} }$.

The scattering process is repeated until the trajectory meets the cloud surface (point {\bf B}).

For a given ${\bf \hat k_{obs}}$ the above process is repeated $M$ times to sample the possible trajectories that intercept the observer along this direction.  A new direction ${\bf \hat k_{obs}}$ is set and the process repeated until $NM$ trajectories have been generated with weights $W_{nm}$.

The mean relative intensity at {\bf A} is then

\begin{equation}
\label{meanint}
\frac{J({\bf A})}{I_0}=\frac{1}{N}\sum_{n}^{N} \frac{I({\bf \hat{k}_{obs}},{\bf A})}{ I_0}=\frac{1}{NM}\sum_{n}^{N} \sum_{m}^{M} W_{nm}.
\end{equation}

The number $N$ of directions is set by the resolution $\mathcal{R}$ of
the model cloud - the optimum $N$ is of the order $\mathcal{R}^2$; adjacent rays propagating in straight lines never separating by more than a cell's width.  In practise 16\% of this optimum number of directions is sufficient to obtain similar results (within 3\%) whilst dramatically reducing the runtime of the calculation.  The number $M$ of samplings of each direction is mainly determined by the cloud's optical properties, $M\sim 10$ being sufficient to obtain satisfactory convergence of the results (within 3\% of those with $M=100$).

%%%%%%%%%%%%%%%%%%%%%%%%%%%%%%  THE FIGURES

\newpage
\begin{figure}
\label{denmaps}
\begin{center}
\plotone{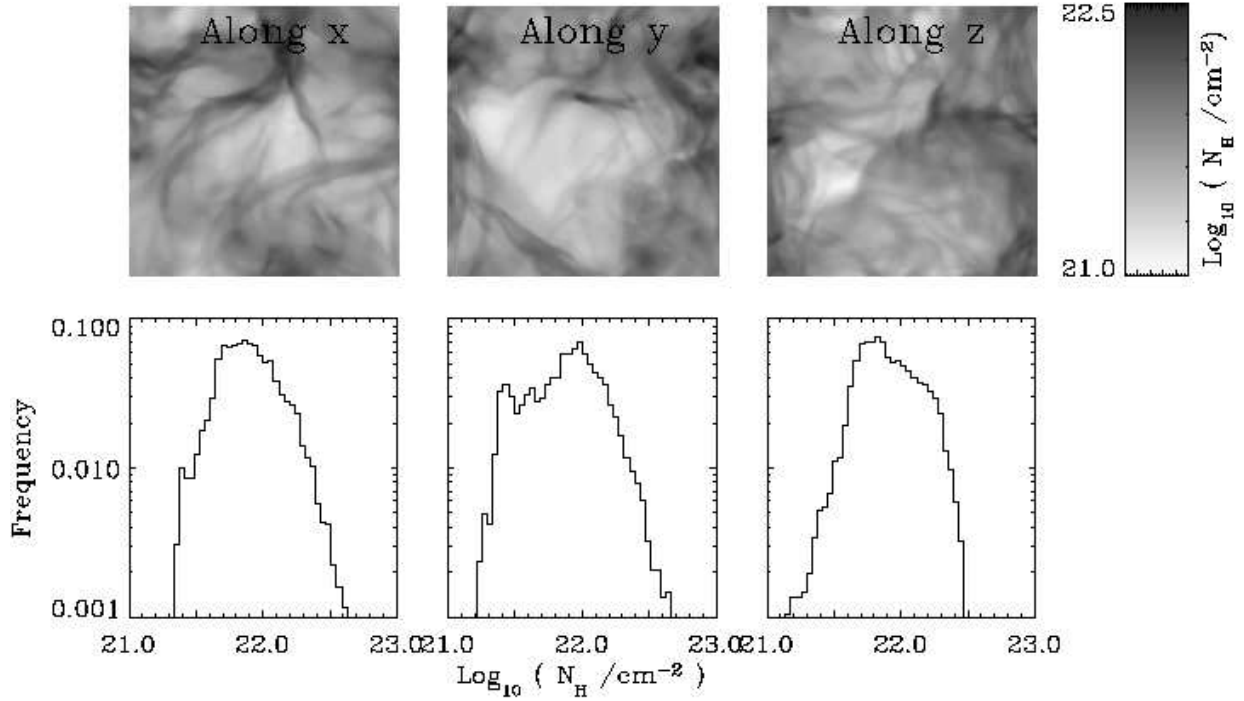}
\end{center}
\caption{Surface density maps for model cloud $A$ looking along the three principal axes (\textit{top row}).  Histograms for the column density $N_H=\int n_H ds$ are shown (\textit{bottom row}) for each viewing direction.  The maps each have a mean column density $<N_H>=10^{22}$ cm$^{-2}$.} 
\end{figure}

\newpage
\begin{figure}[t]
\begin{center}
\plotone{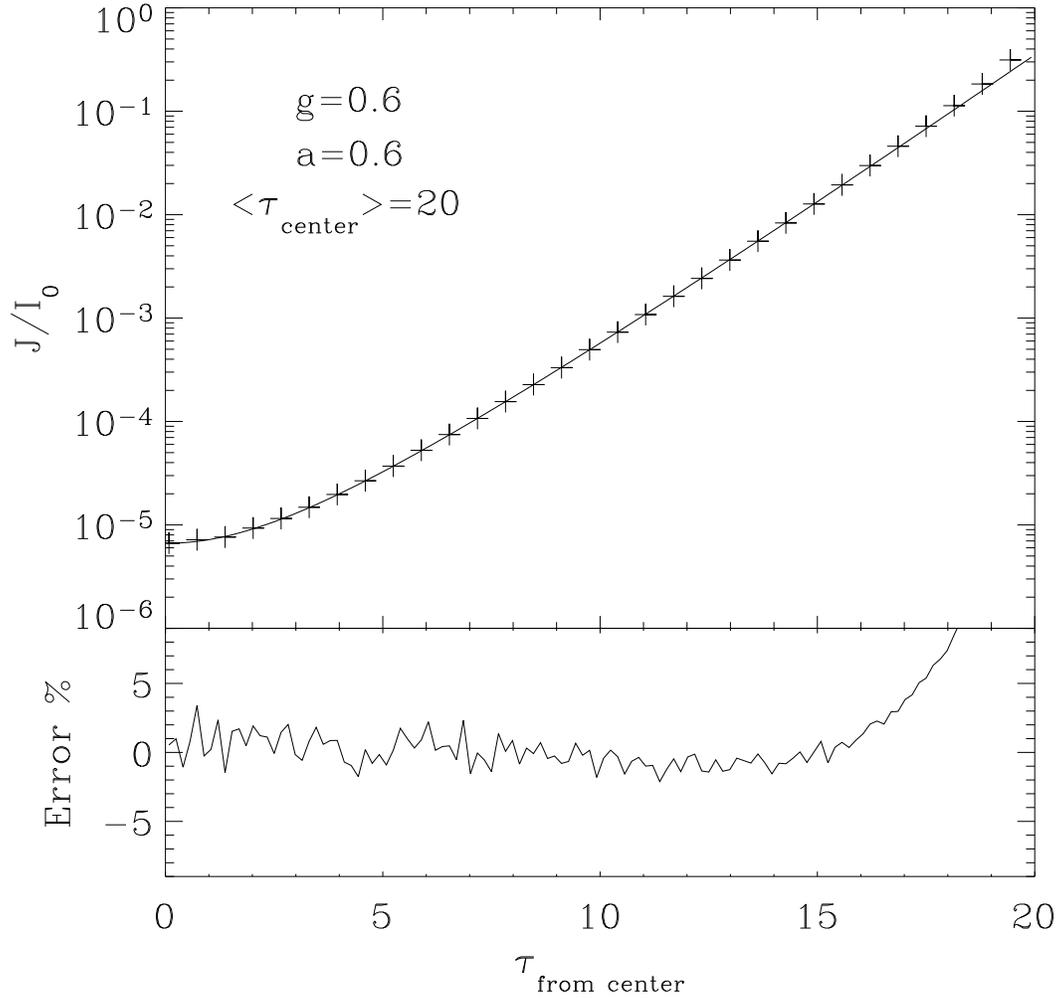}
\end{center}
%\label{test}
\caption{\label{fig:test} The relative mean intensity $J/I_0$ computed with the code (\textit{crosses}) at radial positions $0<r<R$ inside a uniform cloud of central optical depth $\tau=20$ and radius $R$.  The solid line is given by the closed-form asymptotic solution for optically thick clouds derived in Flannery, Roberge \& Rybicki (1980).}
\end{figure}

\newpage
\begin{figure}
\label{fig:res_test}
\begin{center}
\includegraphics[width=15cm]{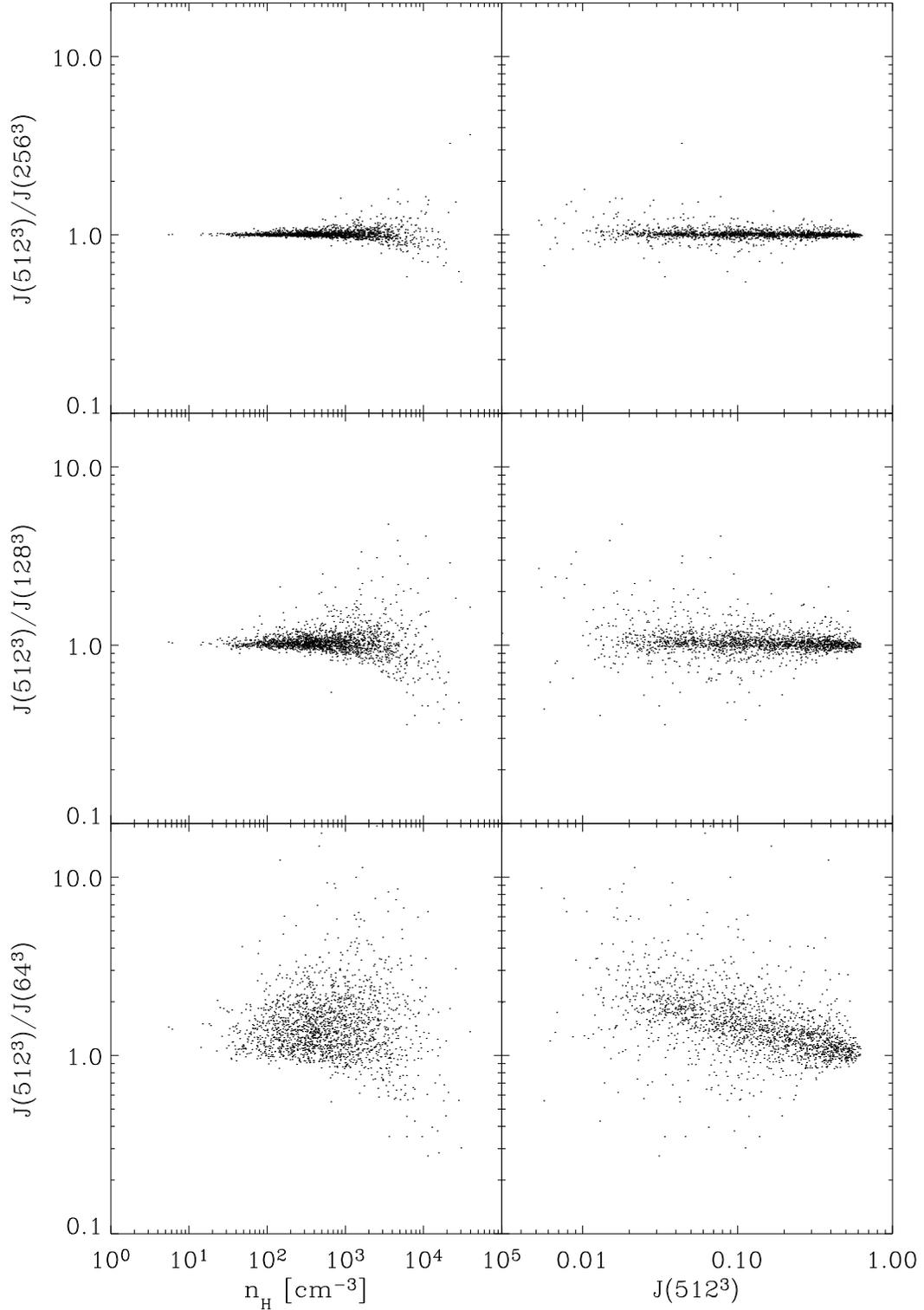}
\end{center}
%\label{res_test}
\caption{Comparison of mean intensity at equivalent points (i.e cell coordinates $(x,y,z)$ 
in the 512$^3$ cell model become $(x/2,y/2,z/2)$ after smoothing to 256$^3$ cells) in model 
\clre \space smoothed from 512$^3$ to 256$^3$,128$^3$ and 64$^3$ cells.}
\end{figure}

\newpage
\begin{figure}
\begin{center}
\includegraphics[width=9cm]{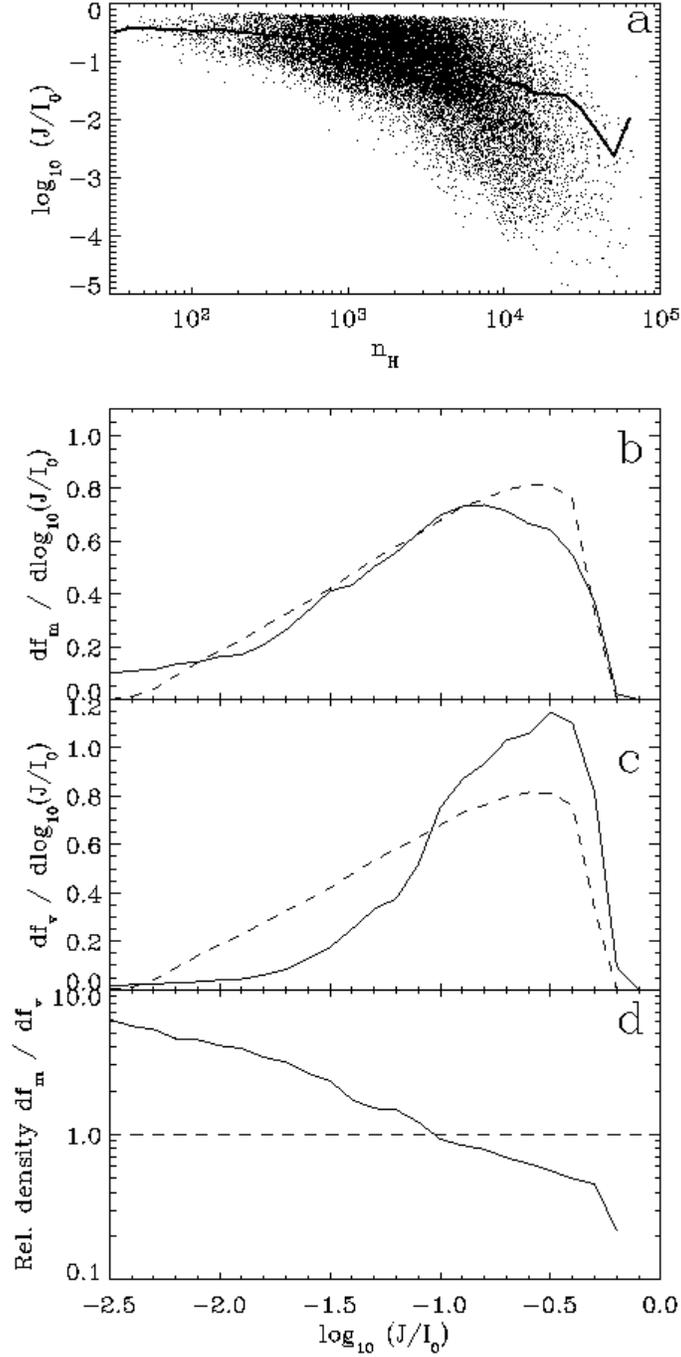}
\end{center}
\caption{\label{fig:bethell_plot}{\bf a} The relative mean intensity \mj \space at $\lambda=550$nm for a random sample of points within cloud $A$, where $I_0$ is the intensity of unattenuated interstellar radiation, assumed to be isotropic. The average \mj \space for a given density $n_H$ is shown by the bold line.  {\bf b} The fraction of the cloud mass per unit $log_{10}($\mj$)$, $df_m/dlog_{10}($\mj$)$. {\bf c} The equivalent volumetric distribution $df_v/dlog_{10}($\mj$)$.  {\bf d} The average overdensity associated with a value of \mj \space (i.e. $df_m/df_v$).}
\end{figure}

\newpage
\begin{figure}
\begin{center}
\plotone{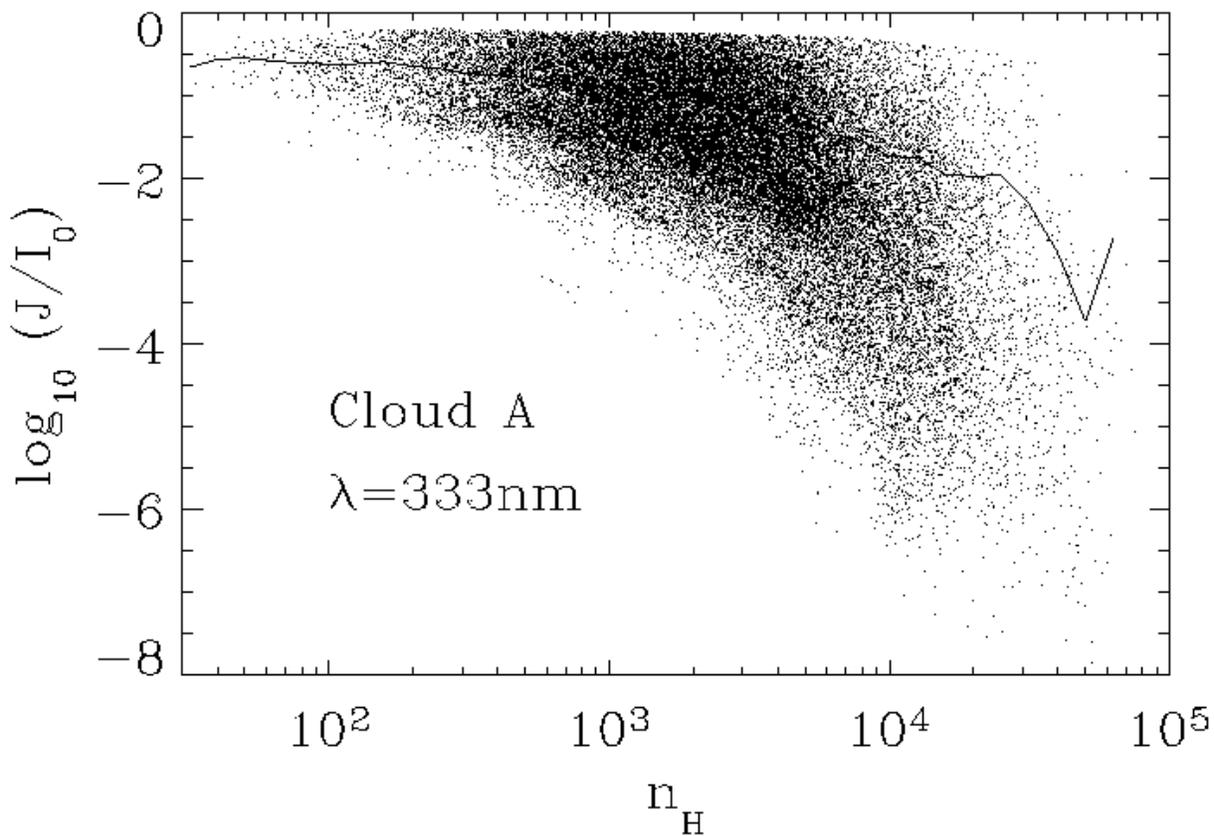}
\end{center}
\caption{Like Figure 4a but calculated at $\lambda=333$nm.  Note the scatter and compare the
scale with Figure 4a.}
\end{figure}

\newpage
\begin{figure}
\begin{center}
\plotone{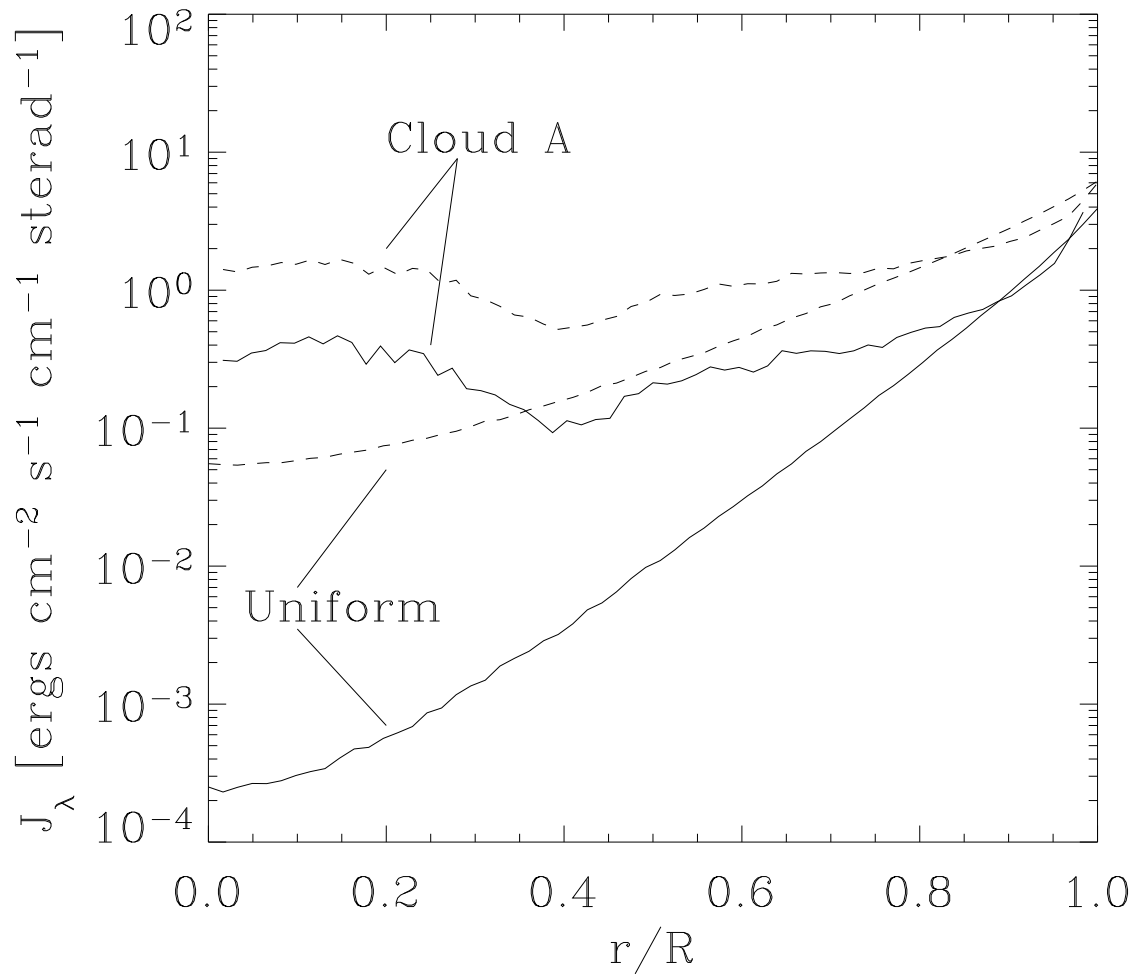}
\end{center}
\caption{\label{fig:UV_onion} For $\lambda=333$nm (\textit{dashed} line) and 
$\lambda=550$nm (\textit{solid} line) the mean specific intensity is averaged over the volume of thin spherical shells to form the shell average $<J_{\lambda}>$.  
The shells are centered on the cloud center, of radii $0<r<R$ and thickness $R/50$, where $R$ is the cloud radius.  
Results for the uniform cloud with the same central optical depth ($<\tau_{cen}>=10$) are shown for comparison.}
\end{figure}

\newpage
\begin{figure}[t]
\label{fig:color_scatter}
\begin{center}
\plotone{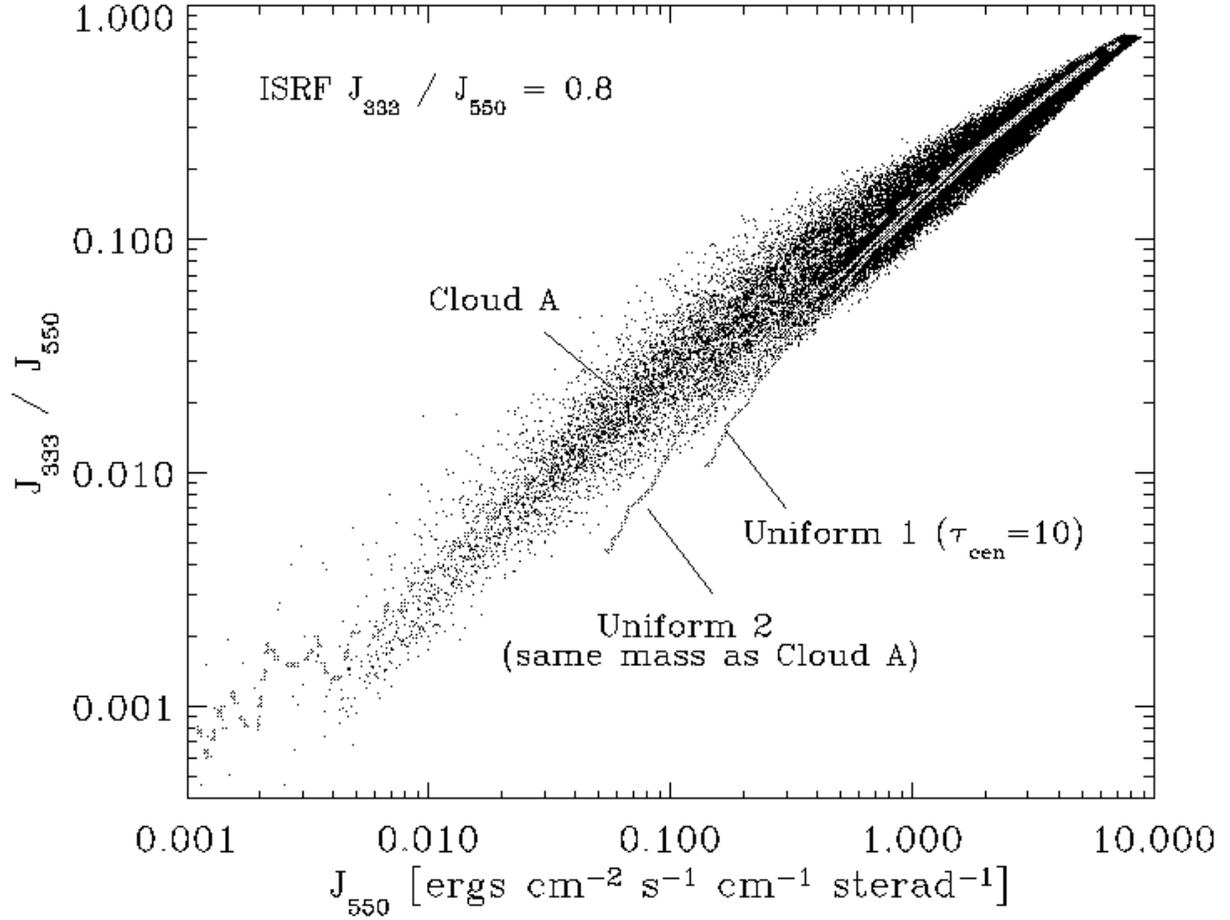}
\end{center}
\caption{The color $J_{333}/J_{550}$ for points inside model \cl.  For comparison the colors for two uniform clouds are shown: Uniform 1 is calibrated to $\tau_{cen}=10$ whereas Uniform 2 has a total mass equal to that of model \cl.}
\end{figure}

\newpage
\begin{figure}
\begin{center}
\plotone{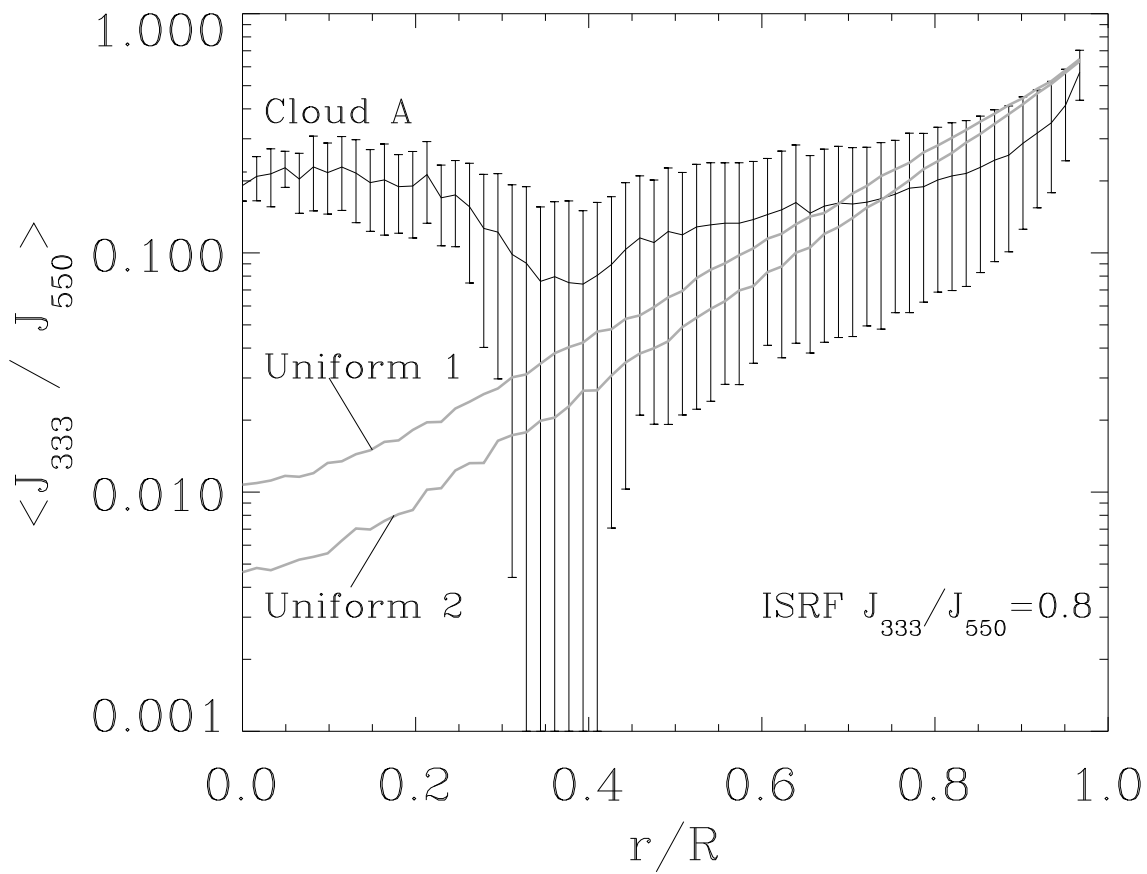}
\end{center}
\caption{\label{fig:color_onion}The color $J_{333}/J_{550}$ averaged in spherical shells 
centered in Cloud \cl.  The averaging is density rather than volume weighted, reflecting 
the average color seen by an H atom or dust grain in these shells.  The error bars represent 
the \textit{rms} scatter about the mean.  For comparison two uniform cloud results are shown; 
Uniform 1 with $<\tau_{cen}=10>$ and Uniform 2 with the same mass as model \cl.}
\end{figure}

\newpage
\begin{figure}
\label{fig:every_bethell_plot}
\begin{center}
\includegraphics[width=14cm]{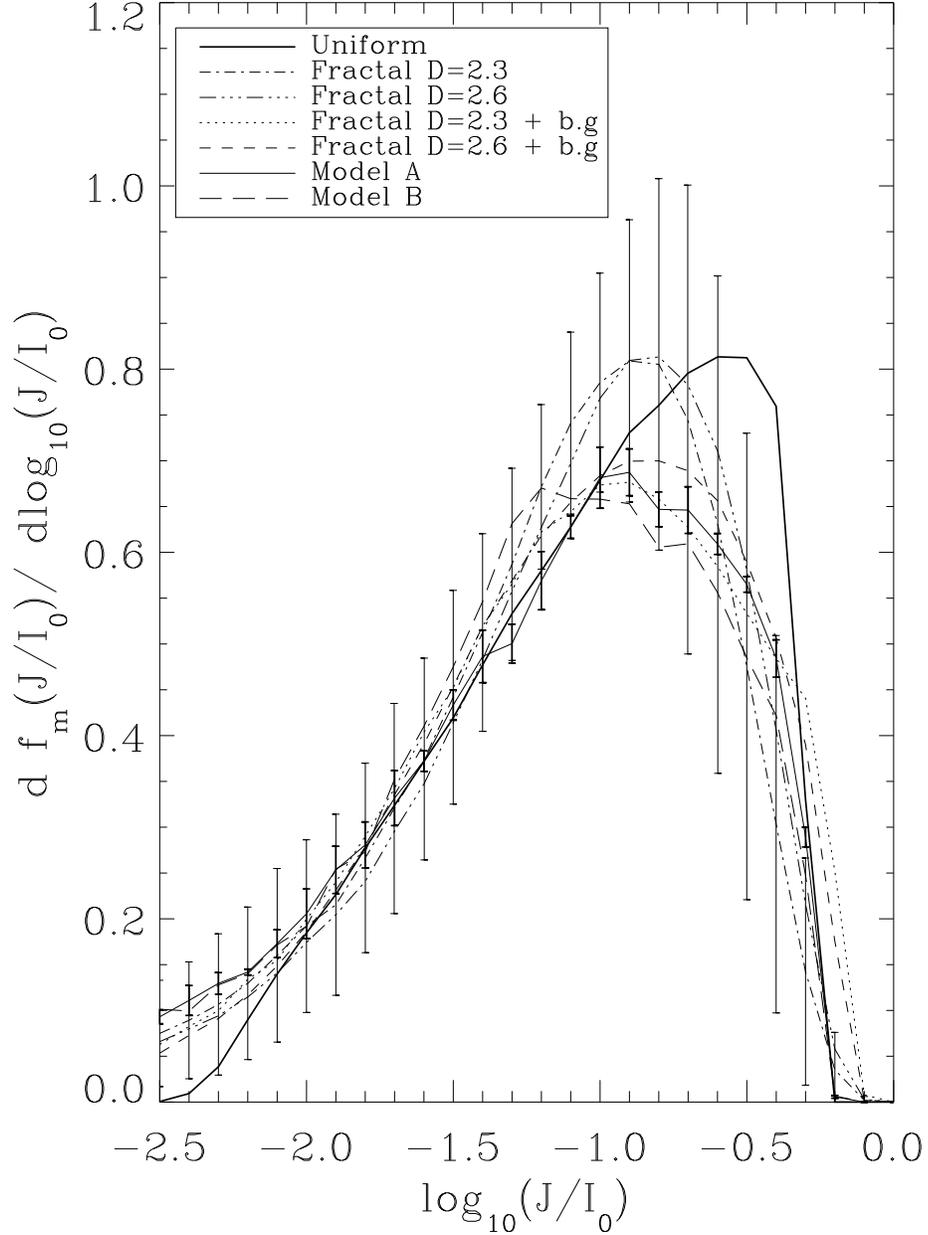}
\end{center}
\caption{Comparison between, uniform, fractal (dimension D, with and without backgrounds `b.g') 
and model clouds \cl \space and \clt .  In the case of the fractal clouds average results have 
been shown with standard deviations (large error bars), generated by averaging f$_m$ for a number 
of clouds with the same physical parameters but grown from different seeds.  The model clouds' 
results are time averages during their steady-state phases (during which there are three time dumps); the deviations (small error bars) 
are then to be interpreted as intrinsic time variations rather than the differences between clouds sharing the same physical parameters but realised separately.}
\end{figure}

\newpage
\begin{figure}
\label{temp_bethell}
\begin{center}
\plotone{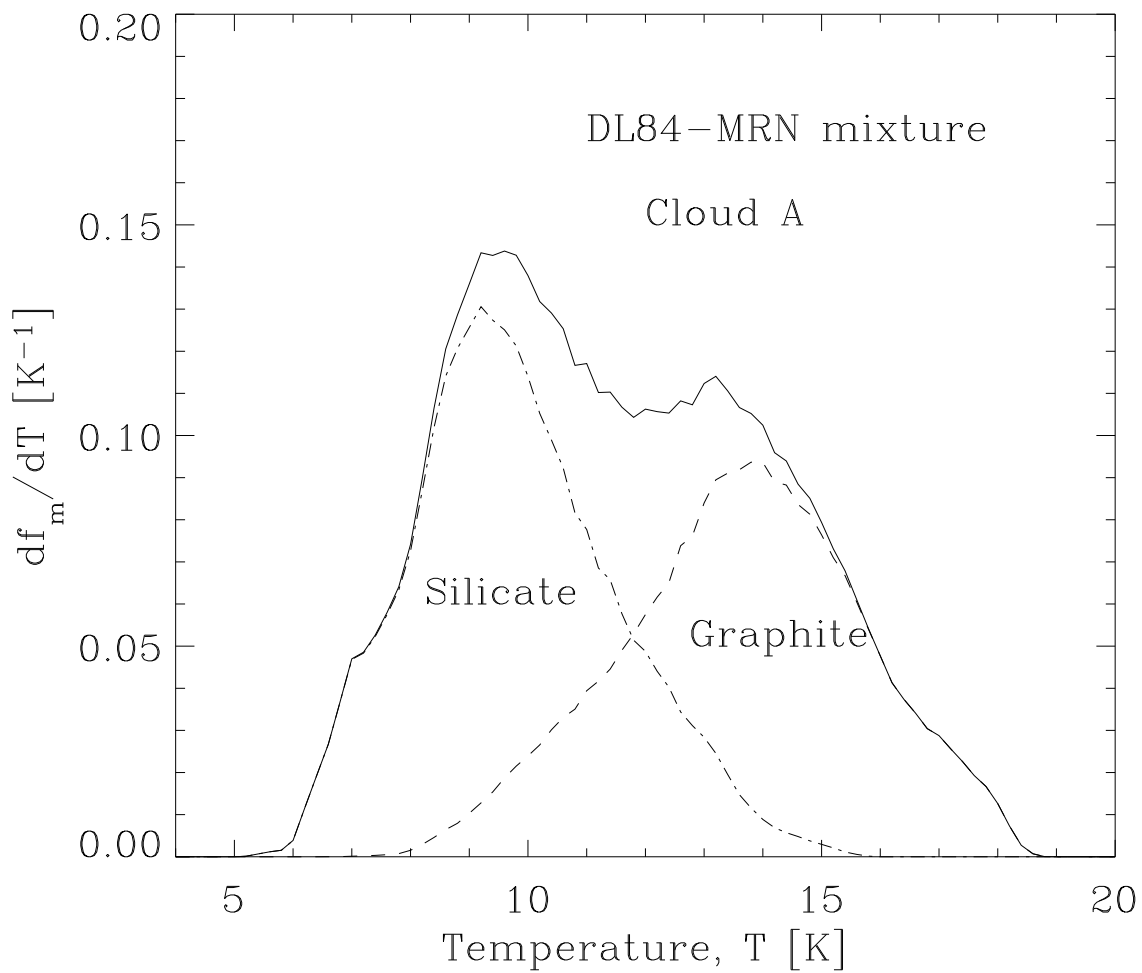}
\end{center}
\caption{The mass distribution function for Cloud $A$ (\textit{solid} line), such that $\int (df_m/dT) dT = 1$.  The mass includes graphite and silicate grains that fall within the MRN size range $[0.005\mu$m$,0.25\mu$m$]$.  The separate contributions of the graphite and silicate grains are shown by the \textit{dashed} and \textit{dot-dashed} lines respectively.}
\end{figure}

\newpage
\begin{figure}
\begin{center}
\includegraphics{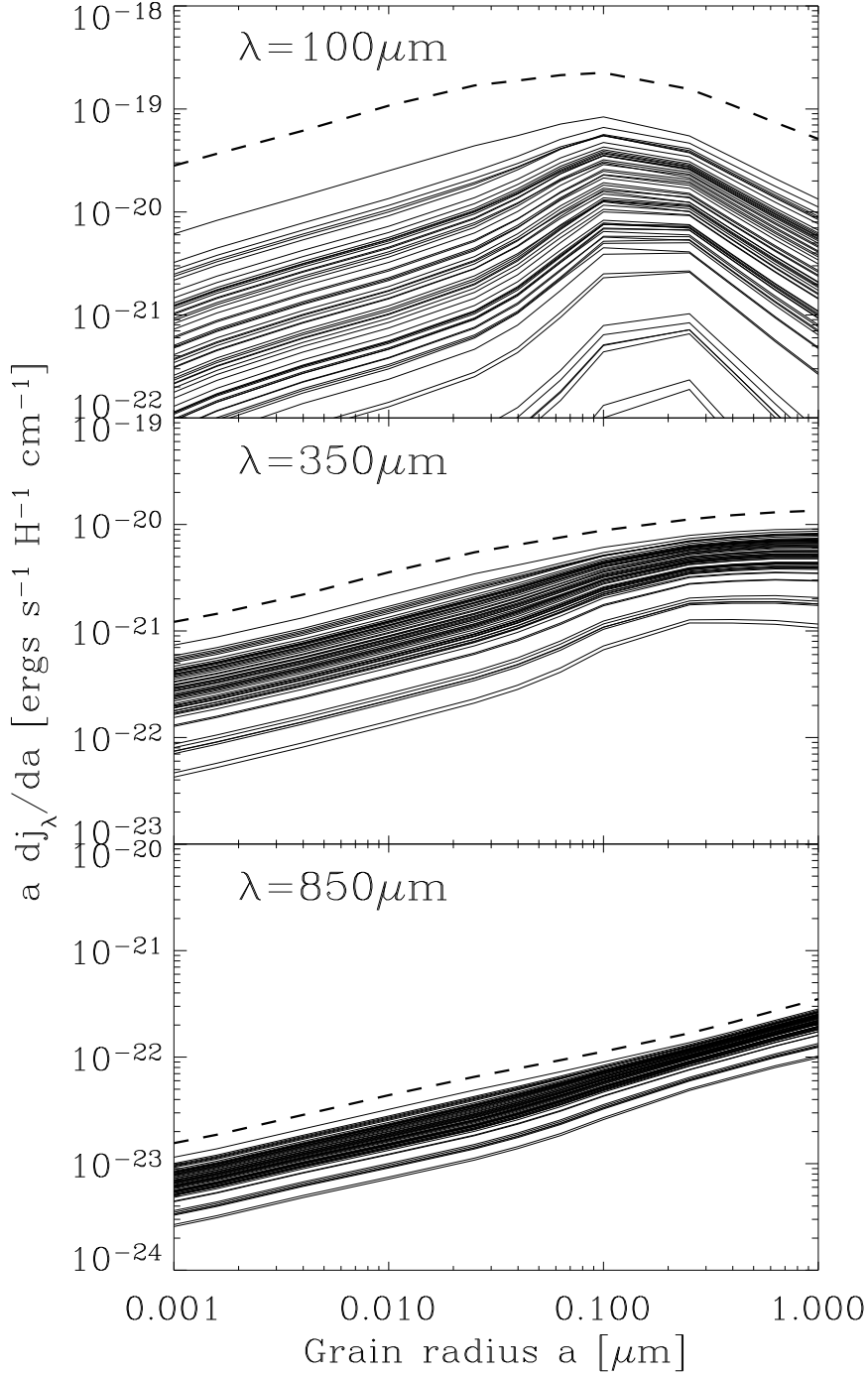}
\end{center}
\caption{\label{fig:emis_by_lambda_by_a}The differential contribution $dj_\lambda /da$ of grains of
  different radii to the spectral
  emissivity $j_\lambda$ at $\lambda=100,350$ and $850$ $\mu$m (\textit{top} to \textit{bottom}).  
  Each line represents the emissivity of dust associated
  with a randomly chosen H atom except the \textit{dashed} line which
  is the emissivity associated with an H atom exposed to the
  \textit{unattenuated} ISRF (i.e. an upper limit).} 
\end{figure}

\newpage
\begin{figure}
\begin{center}
\plotone{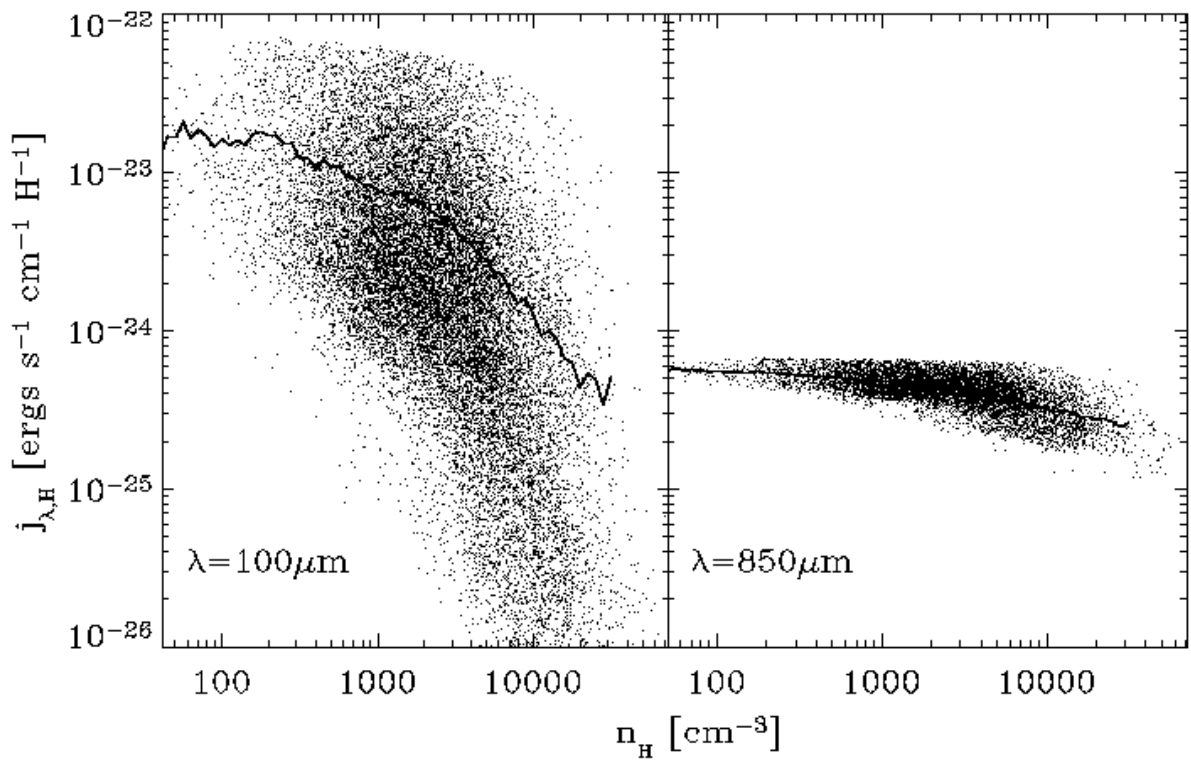}
\end{center}
\caption{\label{fig:emis_scat} Specific emissivity per H atom $j_{\lambda,H}$ 
at $\lambda=100,850\mu$m for a random sample of atoms situated in locations of 
density $n_H$ in model \cl.  The solid lines are the mean values of $j_{\lambda,H}$ 
associated with density $n_H$. }
\end{figure}

\newpage
\begin{figure}
\begin{center}
\plotone{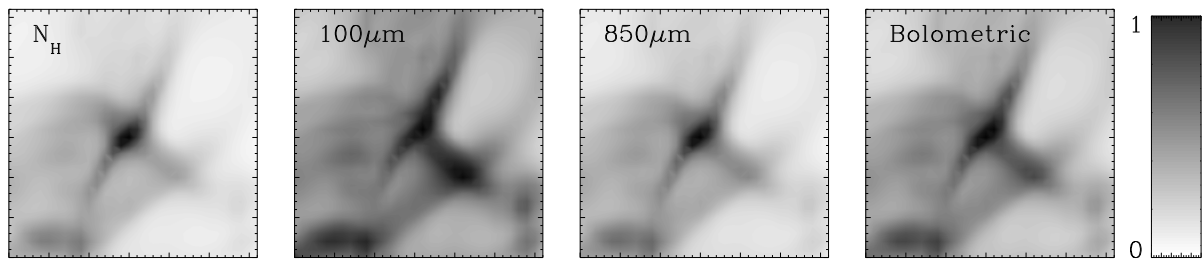}
\end{center}
\caption{\label{fig:brightness_maps} Surface density $N_H$, 100$\mu$m, 850 $\mu$m and bolometric brightness maps (\textit{left} to \textit{right}), scaled such that their maxima equal one.} 
\end{figure}

\newpage
\begin{figure}
\begin{center}
\plotone{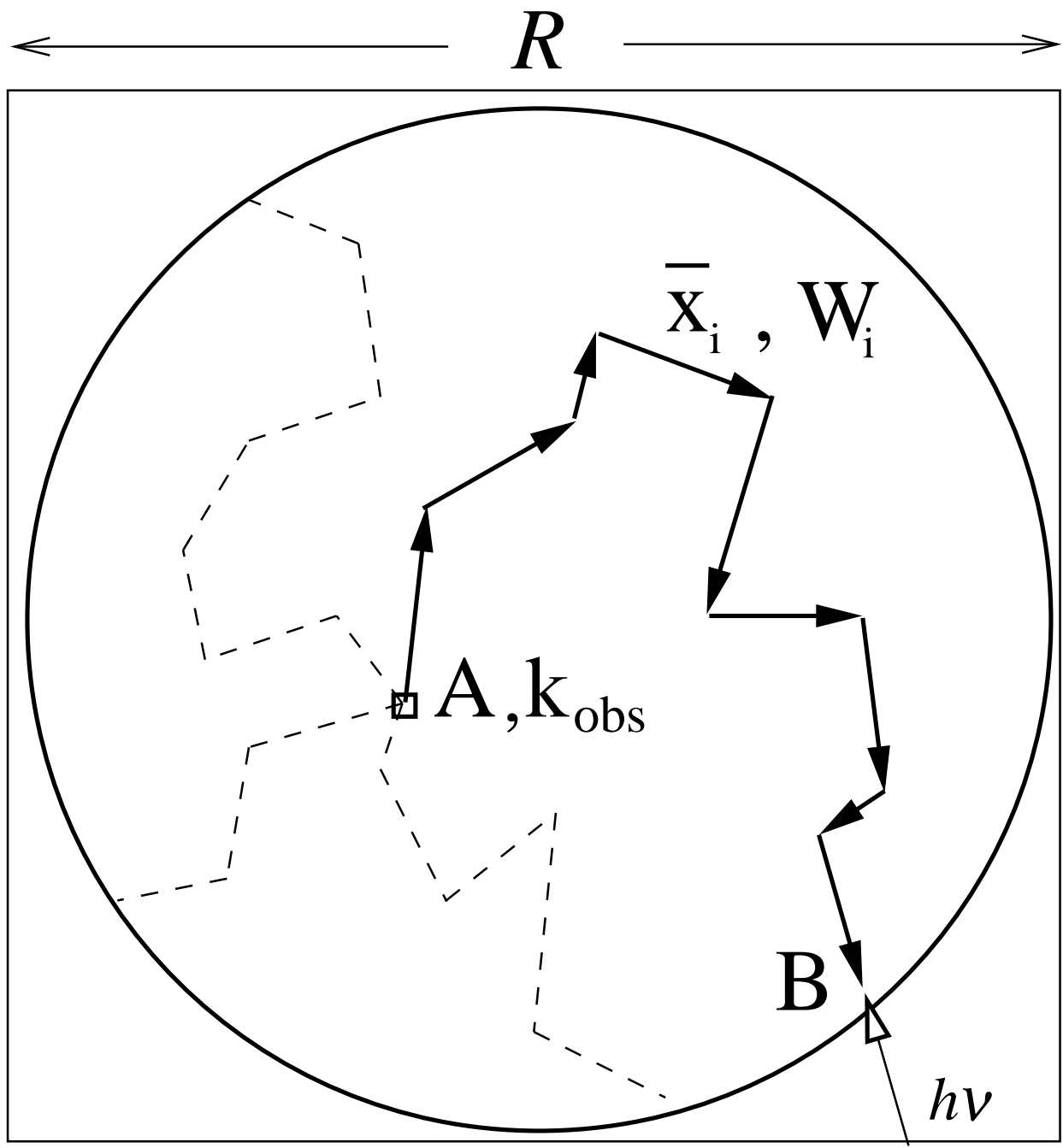}
\end{center}
\caption{\label{diagram} The spherical cloud sits snugly inside the cubic domain of $R^3$ cells.  An observer {\bf A} sends out $M$ rays (one of which is shown) for each of the $N$ directions (${\bf \hat k_{obs}}$) sampling the observe's entire sky. Each ray is evolved according to the Monte Carlo sampling of free paths between scatterings and scattering angles until it reaches the cloud's edge ({\bf B}).  The \textit{dashed} lines are other trajectories joining the observer {\bf A} of photons ($h\nu$) to the ISRF.}
\end{figure}

%%%%%%%%%%%%%%%%%%%%%%%%% THE TABLES

\begin{deluxetable}{cccc}                                                    
\tablewidth{0pc}
\tablecaption{Model list \label{tab:models}}                               
\tablehead{
\colhead{Name}&\colhead{Resolution}&\colhead{$\beta$}&\colhead{Source}}     
\startdata                                                                      

$A$& $128^3$ & $4.04$ &  Heitsch, Mac Low \& Klessen (2001) \\
$B$& $128^3$ & $0.05$ &  Heitsch et al (2001) \\
$C$& $512^3$ & $4.04$ &  Li, Norman \& Mac Low (2004) \\ 

\enddata 
\tablecomments{The MHD simulations used to make Clouds A,B and C.  The parameter $\beta=P_{th}/P_{mag}=8\pi c_s^2 \rho/B^2$.}

\end{deluxetable}                                                              

\begin{deluxetable}{ccccccc}
\label{table_temps}
\tablecolumns{6}
\tablewidth{0pc}
\tablecaption{Average Grain Temperatures T and \textit{r.m.s} scatter $\sigma$}
\tablehead{
\colhead{} &\colhead{} & \multicolumn{4}{c}{Log$_{10}(a/\mu$m$)$} \\
\cline{3-6}
\colhead{ }&\colhead{ }&\colhead{-3.0}&\colhead{-2.0}&\colhead{-1.0}&\colhead{0.0}} 
\startdata

Graphite   &   T (K)   &   12.4   &   13.1   &   14.5   &    10.12\nl
   &   $\sigma$   &   2.1   &   2.2   &   1.9   &   1.1\nl
\cline{1-6}
Silicate    &   T (K)  &   9.5   &   9.6   &    9.8   &   10.2\nl
   &   $\sigma$   &   1.8   &   1.9   &   1.6   &   1.1\nl
\enddata
%\caption{The average temperature }
\end{deluxetable}

\end{document}